# PRODUCTIVITY ENHANCEMENT THROUGH PRODUCTION MONITORING SYSTEM

**Shiva Prasad H C[1], Potti Srinivasa Rao[2], Gopalkrishna B[3] and Aakash Ahluwalia[4]**

***Abstract:*** *A production monitoring system uses the real time data while production is online. The real time production monitoring systems are designed as means of auto data to collection and monitoring the data via display boards. This study focuses on analysing the real time production monitoring systems through trend analysis in production and over consumption of raw material controlling the over consumptions in a pen manufacturing industry. Methodology followed is through process flow diagram, collection of data, analysis of data. Pre and post analysis was conducted to identify the factors responsible for over consumption and causal factors responsible for the over consumption were identified to reduce the cost of consumption by 58% with introduction of production monitoring system.*

## I. INTRODUCTION

Productivity is an important factor for achieving high efficiency and accuracy in the production line with the proper utilization of available resources. Precise data management at the shop floor and monitoring systems are equally important for improving production performance. Many production processes are following the manual methods of collecting data at the shop floor leads to inconsistencies and inaccuracies. In a competitive environment, a tool for operation manager is required on the shop floor to create a well-organized database on the shop floor and to offer a better solution to a medium sized production companies to attain productivity. During many occasions the human interventions in the process of collecting the data, the data is considered to be less reliable (Subramaniam *et al.*, 2009). This research work focuses on the Production Monitoring System (PMS) as applied to a Pen Manufacturing Industry thus analysing the data and making decisions for

---

[1,2,3] Department of Humanities and Management, Manipal Institute of Technology, Manipal University, Manipal, India, *E-mails* [1]*hcs.prasad@manipal.edu;* [2]*srinivas.potty@manipal.edu;* [3]*gopalkrishna.b@manipal.edu*

[4] PG Scholar, Department of Humanities and Management, MIT Manipal University, Manipal, India, *E-mail:* [4]*aakash.ahluwalia@gmail.com*

improving the quality and quantity in order to incorporate suitable changes in the production line.

*Production Monitoring System* is a tool used by management to gather and distribute information about the scenario of the shop floor. It basically helps in achieving goals by reducing the down time and by increasing productivity (Snatkin, Karjust, and Eiskop, 2012). The objective of any PMS is designed around to collect and distribute real time data from the shop floor. The collected data may necessitate towards decision making. The online data is monitored during the production process and responding immediately in a proactive method so as that the end results obtained (Phaithoonbuathong et al., 2010). Such system acts as an alarm and warns the respective department concerning in recognizing the process fault. A PMS display the current production data with an additional facility of reporting and administrating the production module (Gourgand, Lacomme and Traoré, 2003). The stored data is then analysed to detect the production output, estimations, projections for quick decision making and production planning. The faults are detected that must decrease the wastage of time in maintenance and improve overall equipment effectiveness (Snatkin, Karjust and Eiskop, 2012). Customers' focuses on zero defect products, quality and expect the job to be delivered on time. This is achieved by eliminating the process variability that causes defects and they need to recognize immediately when and where there are problems and take preventive measure to avoid the possible defective product being delivered to the customer. Wholesalers and retailers are more concerned about the quality. In case of any problem, their reporting and feedback inputs are important for the organization to access the reason for the problem in the production line. This feedback also helps the organization to improve on the production line and hence to improve the quality and profitability (Sharma, 2010). In ever increasing global economy the productivity enhancement is achieved by maintaining a competitive edge over the competitor. The tools such as Kaizen, Lean Manufacturing, Six Sigma, Total Quality Management and Continuous Improvement are implemented in many manufacturing units to edge the operation processes. Though the manufacturing units have formalized the concept of continuous improvement initiatives through the visual management techniques to gain efficiency lacks in the learning process in the field the PMS, Continuous Monitoring System (CMS) and amongst all the available tools, PMS is most popular as it pin points the factor of production (Sharma, 2010). By examining the causal factors responsible for the over/under consumption and recommends issues so as overcome the effects the production.

The objectives of this research is to make a comparative analysis of consumption based on moulding average weights with respect to standard weights before and after the examination of factors those effects. To conduct the cost incurred profit and loss analysis in terms of production and consumption and to detect the factors of over consumption (Lei, and Chan, 2012). Finally aiming to minimizing the influence of factors of consumption and achieve the goal.

The present scenario is that level production at low cost is the top priority for every Industry and in a Plastic Industry, the issue over consumption is always high thus affecting the production at every end. In this project work various factors that influences the over consumption and the possible outcomes were derived to nail down the problem of over consumption.

## II. METHODOLOGY

The study is carried out in Pellow Plastic Products Limited (pseudo company name), India. The company manufactures different types of writing pens. The company manufactures different plastic products such as household products and writing instruments (pens). The data was collected from June to March. This research focuses on the writing pen instrument manufacturing division. Shop floor consists of 24 injection moulding machines. Monthly production data were collected from each injection moulding machine displayed on their I-bed Control panel by using industrial laptop and further the data is stored in the form of excel sheets.

### Procedure of collecting the data

* *Production data collection-* Data is collected on the basis of monthly target that gives the good count of production and rejection. The collected data gives the amount of total raw material consumed for the production. The collected data is analyzed on the basis of consumption of raw material and then the number of products (pens) is calculated according to the standard weights of different products. Analyzing the data gives out the factors responsible for over consumption. The data was further critically analyzed for the causal factors responsible for the over consumption of raw material.

* *Production data analysis-* Analysis of data is based on the consumption of raw material for the actual production carried out in the industry and the standard consumption of raw material. Actual consumption is the amount of raw material consumed for the actual production whereas standard consumption is the amount that is supposed to be consumed for fulfilling the monthly target.

* *Analysis output-* It is based on the cost analysis and Profit/Loss analysis in terms of consumption and production.

- *Factors responsible-* Further analysis brings out the factors responsible for varying trend in the production as well as consumption of raw material.

- *Optimum solution-* Gives out the best possible solution to overcome the factors responsible for variation in production and consumption.

Process flow diagram shows the step-by-step methodology flow (Fig. 1) that helps in analysis of process parameters. Initially the production data was collected on the basis of total consumption of raw material. After the collection of data, analysis was carried out on the basis of variance between the actual consumption and standard consumption. The data was further critically analysed to see the production trend and to examine the causal factors responsible for over consumption and finally measures were taken to overcome those factors.

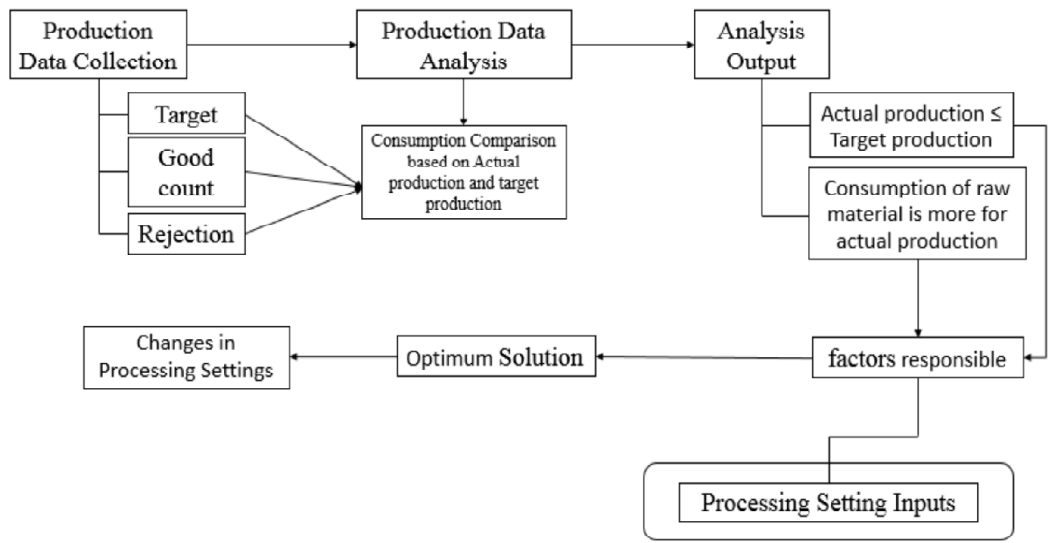

**Figure 1: Process Flow for Methodology (Source: Raharno and Martawirya, 2011)**

**Data Collection:** Production data was collected for 12 months, divided into two phase. In the pre-phase data is collected for six months and post-phase the data collected for another six months. In pre-phase data is analysed for various factors of over consumption and in post-phase two measures were taken to overcome the factors responsible for over consumption.

The comparative analysis was made on the basis of actual consumption of raw material and standard consumption of raw material. Actual consumption is the amount of raw material consumed during the production. Standard consumption of raw material is an estimated amount of raw material supposed to be consumed for a particular amount of production.

## III. RESULTS AND DISCUSSION

### Comparative analysis of pre-phase and post-phase raw material consumption

A Comparative analysis was made on the basis of data collected for the actual consumption and standard consumption of raw material during pre-phase. Variance on the basis of standard consumption and actual consumption is calculated and hence the cost analysis is carried depending upon the total over consumption of the raw material (Table 1).

**Table 1**
**Consumption of raw material in pre-phase**

| Months | Standard Consumption (Kgs) | Actual Consumption (Kgs) | Variance (Kgs) | Avg. Cost @ INR 150/Kg/year |
|--------|----------------------------|--------------------------|----------------|------------------------------|
| April | 82162 | 84063 | 1901 | 2,85,150 |
| May | 52278 | 54388 | 2110 | 3,16,500 |
| June | 118981 | 121243 | 2262 | 3,39,300 |
| July | 97382 | 100194 | 2812 | 4,21,800 |
| August | 97987 | 100442 | 2455 | 3,68,250 |
| September | 85454 | 87789 | 2335 | 3,50,250 |
| Total | 534244 | 548119 | 13875 | 20,81,250 |

*Note:* Assumed average cost of polymer used for pens is INR 150/kg/year

The amount of raw material actually consumed during pre-phase and its variance in terms of weight shows that spike in July month (2812 Kgs) with a valley in April month (1901 Kgs.) (Table 1). It is observe, actual consumption is more as compared to the standard consumption. According to the production scenario it is observed that the actual production is almost equal to the target production, but the actual consumption is higher than the standard consumption. The over consumption for six months is 13,875 kgs, the cost incurred for the raw polymer is INR 2,081,250 at the rate of INR 150/kg/year.

### III. (A) Profit/Loss Analysis

On the basis of average weight of different products (pens) it is calculated that 13,875 kgs. of raw material is enough to increase the monthly production by 2.5-4.5% and profit is 3%.

Combined average weight of the raw materials of all the pens is: (0.00675+0.00723+0.00782+0.00831)/4=0.0075275 kgs.

**Table 2**
**Average weights of different products (pens)**

| Product Name (Pen Name) | Average weights (in kgs) |
|---|---|
| Fine Grip | 0.00675 |
| Gripper | 0.00723 |
| Techno Tip | 0.00782 |
| Pin Point | 0.00831 |
| Combined Average Weight | 0.0075275 |

The total production and profit is calculated based on the total over consumption (Table 1) with the combined average weight (Table 2)

Production loss during Pre- phase: 13,875/0.0075275 = 1,843,241 pens

Average cost of production of one Pen = INR 2.65

Average market cost of one Pen = INR 6.00

Profit/Pen = INR 3.35 (6.00-2.65)

Total loss (Phase 1) = 1,843,241 x 3.35 = INR 6,174,858

Average estimated monthly loss = INR 6,175,858/6 = INR 1,029,143/month

The production and loss out of that production is basically the hidden loss suffered by the industry every month. There are various factors responsible for this loss.

### III (B) Factors Responsible

One of the factors that are responsible for over consumption is '*Production Processing/Programming settings*'. Production processing is the machine coding in that various parameters are taken into account such as processing temperature, injection speed, injection pressure, injection time and various inputs are given on the basis of different properties of the raw polymer. It depends on:

*Processing Temperature (PT) factor-* It is the temperature at that the polymer melts and is ideal for processing. It varies from material to material used for the pen.

*Injection Speed (IS) factor-* It is the speed at that the material is injected from the injection unit into the mould so as to fill the mould completely. It is less for small size products and high for large size products.

*Injection Pressure (IP) factor-* It is the pressure at that the material is injected from the injection unit into the mould so as to fill the mould completely.

*Injection Time (IJ) factor-* The time for that the material is injected into the mould that fills the mould cavity completely.

*Cooling Time (CT) factor-* It is the time taken by the mould to cool down the material in it so as to avoid various defects such as shrinkage, blow marks. Cooling is basically done by running water through the cores of the mould. Production process settings are considered to be the important factor responsible for the over consumption because: Even a slight variation in any of the listed five parameter can change the weight of the product. It can create small defects that were visual inspected.

Changes in processing temperature, injection speed, injection pressure, injection time and cooling time is required as per the properties and requirement of different polymers such as Polypropylene, Polycarbonate, Polystyrene and other polymers. Processing standards of these materials are given below (Table 3).

**Table 3**
**Properties of Polypropylene, Polycarbonate and Polystyrene**

| Polymers | Density | Processing Temp. |
|----------|---------|------------------|
| Polypropylene | 0.91-0.93 g/cm$^2$ | 220-250º C |
| Polycarbonate | 1.2 g/cm$^2$ | 230-280º C |
| Polystyrene | 1.05 g/cm$^2$ | 280-320º C |

*Source:* Injection Moulding Data book by Demag Plastic Group, 2004

Injection speed, injection pressure, injection time and cooling time varies from product to product and for the mould to mould. The over consumption of raw material is because of the variation in the processing settings. During the pre-phase during months of April to September, machines were made to work with these parametric settings (Table 4). It is seen that the productivity was almost higher with the target but the only problem faced was the variation in the weight of the product. As an individual product, variation is negligible, but in a mass production the variation came out to be very large. The controlled variables for production processing are manipulated problem of over consumption and are being addressed by changing the production processing settings by taking number of trials for different raw material used (Table 4).

The critical factor is the in the processing temperature, slight incremental processing temperature make the material more viscous and hence the inflow of the material is high, thus leading to incremental in weights.

Total 16 trails were taken by considering the following assumptions:

**Table 4**
**Processing Settings for Pre-phase**

*Polycarbonate and Polystyrene*

| Processing Settings | Polymer-Polypropylene | Polymer-Polycarbonate | Polymer-Polystyrene |
|---|---|---|---|
| Processing Temp. | 280º C | 295º C | 340º C |
| Injection Speed | 75 mm/sec | 130 mm/sec | 90 mm/sec |
| Injection Pressure | 175 bar | 143 bar | 120 bar |
| Injection time | 11 sec | 8.7 sec | 07 sec |
| Cooling Time | 06 sec | 11-13 sec | 08 sec |

1. Temperature is kept constant so as to maintain the viscosity of the melted material. If the viscosity changes, the inflow of the material will be less or more, depending on the temperature.

2. Speed and pressure are maintained in such ways that the product produced are defect free. There are number of defects that are observed during the trials. Major defects are shrinkage, short piece inside the mould due to less inflow of material, over packing of mould due to more inflow of material, silver marks on the product due to low temperature.

3. Trials were taken in order to obtain a perfect product with accurate standard weight.

4. New production processing settings are obtained by taking number of trials, sample trials are shown in Table 5 along with the processing temperature, injection speed, injection pressure, type of defects seen during the trials and weight of the product.

It is observed that by keeping the temperature constant at 245ºC and by varying the speed and pressure, new processing parameters for Polypropylene (Table 5) are obtained that are: Injection Speed: 130 mm/sec; Injection Pressure: 145 bar; Defects: No defect; Weight: Equivalent to standard weight. The product was accepted.

Observation on the change made with the settings listed gives defects like short piece, shrinkage and over packing. Similarly, trials taken for polycarbonate and polystyrene are given in Table 6 for the optimized results.

The parameters that are for the new processing parameters are derived by taking numbers of trial (Table 6). For post processing settings, consumption is collected on the basis of the given production processing settings.

**Table 5**
**Sample Trials for Polypropylene for new processing settings**

Trials for Polypropylene
Assumption: Temperature is kept constant

| Trial No | Temp. $^0$C | Speedmm/sec | Pressurebar | Defects | Weight | Remark |
|---|---|---|---|---|---|---|
| 1 | 245 | 90 | 100 | short Piece | - | reject |
| 2 | 245 | 95 | 105 | short Piece | - | reject |
| 3 | 245 | 100 | 110 | short Piece | - | reject |
| 4 | 245 | 105 | 115 | short Piece | - | reject |
| 5 | 245 | 110 | 120 | short Piece | - | reject |
| 6 | 245 | 115 | 125 | short Piece | - | reject |
| 7 | 245 | 120 | 130 | short Piece | - | reject |
| 8 | 245 | 125 | 135 | minor cut on the edges | - | reject |
| 9 | 245 | 127 | 138 | minor cut on the edges | - | reject |
| 10 | 245 | 129 | 143 | minor cut on the edges | - | reject |
| 11 | 245 | 130 | 145 | No defects | standard | accept |
| 12 | 245 | 135 | 150 | shrinkage over packing of mould | excess | reject |
| 13 | 245 | 140 | 152 | shrinkage flow mark over packing of mould | excess | reject |
| 14 | 245 | 143 | 155 | shrinkage flow mark over packing of mould | excess | reject |
| 15 | 245 | 146 | 158 | shrinkage flow mark over packing of mould | excess | reject |
| 16 | 245 | 150 | 160 | shrinkage flow mark over packing of mould | excess | reject |

**Table 6**
**The post-phase processing settings for Polycarbonate and Polystyrene**

| Processing Settings | Polymer-Polypropylene | Polymer-Polycarbonate | Polymer-Polystyrene |
|---|---|---|---|
| Processing temp. | 245$^o$ C | 230$^o$ C | 310$^o$ C |
| Injection speed | 130 mm/sec | 155 mm/sec | 120 mm/sec |
| Injection pressure | 145 bar | 120 bar | 90 bar |
| Injection time | 07 sec | 06 sec | 06 sec |
| Cooling time | 05 sec | 08 sec | 4.8 sec |

**Table 7**
**Consumption of raw material in post-phase**

| Month | Standard Consumption (Kgs) | Actual Consumption (Kgs) | Variance (Kgs) | Cost INR 150/kg |
|-------|---------------------------|--------------------------|----------------|-----------------|
| Oct | 89,554 | 90,839 | 1,285 | 192,750 |
| November | 69,099 | 70,213 | 1,114 | 167,100 |
| December | 93,202 | 94,192 | 990 | 148,500 |
| January | 107,581 | 108,465 | 884 | 132,600 |
| February | 117,026 | 117,773 | 747 | 112,050 |
| March | 101,051 | 101,840 | 789 | 118,350 |
| Total | 577,513 | 583,322 | 5,809 | 871,350 |

On the basis of total over consumption (Table 7) and combined average weight calculated (Table 2), total production and loss is calculated for the post phase.

Loss of Production – 5,809/0.0075275 = 771,703 Pens; Average Cost of Production of one pen = INR 2.65

Average Market Cost of one pen = INR 6.00; Profit/Pen = INR 3.35; Total Loss = 771,703 x 3.35 = INR 2,585,205

Average estimated monthly profit = INR 2,585,205/6 = INR 430,867/Month

The comparison indictor gives in-terms of quality output based on pre-process and post setting of production process (refer Fig. 2). Over consumption is reduced to 5,809 kgs, production loss in terms of pieces is reduced to 771,703 pieces and loss in terms of cost is reduced to INR 2,585,205 for six months and it is observed that over consumption is reduced by 58% (Table 2 and Table 7).

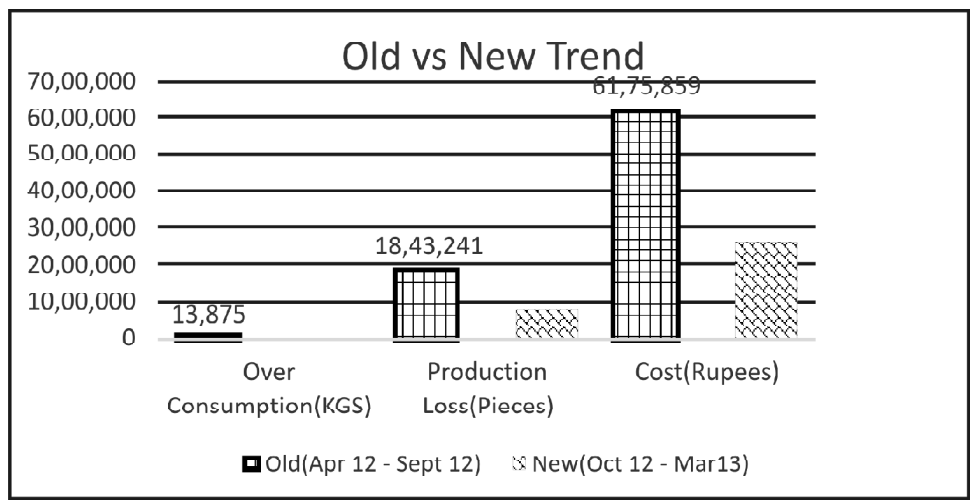

**Figure 2: Comparison of pre-phase against post-phase**

### III. (C) Factor Validation

The process is monitored using a control and helps to discover the process variability. Typically control charts are used for time-series data, though they are used for validating the process meeting the designed specification data that is logically comparable. Validation of results obtained by machine manipulating variables is tested using control chart as process control. Thus the values obtained for the production during post-phase are well within the limit of obtaining a standard weight quality product.

Analysis of the control chart indicates that the process is not a chance cause but in long run it was under control and is stable, with little variation that is common in the process. It is observed that no necessary corrections in the process control parameters are required. The data from the process is suitable for the predicting the future process performance. In the chart (Fig. 3) it is evident that the monitored process is well within the control limit and the analysis of the chart helps to determine the sources of variation.

A process that is stable but operating outside of desired limits (e.g., scrap rates may be in statistical control but beyond desired limits) needs to be improved through a deliberate effort to understand the causes of current performance and fundamentally improve the process (see Table 8) through outcome of weight by giving various inputs in term of temperature, speed and pressure.

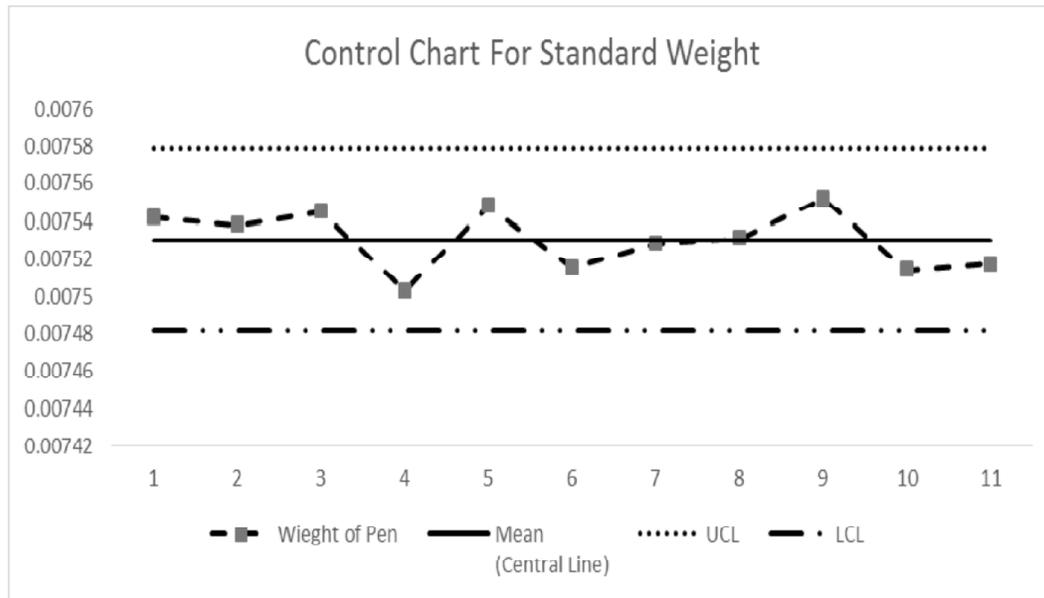

**Figure 3: Control Chart**

**Table 8**
**Process control chart for each pen weight**

| Weight of pen | Mean (Central Line) | UCL | LCL |
|---|---|---|---|
| 0.007542 | 0.007530273 | 0.007579 | 0.007482 |
| 0.007538 | 0.007530273 | 0.007579 | 0.007482 |
| 0.007545 | 0.007530273 | 0.007579 | 0.007482 |
| 0.007503 | 0.007530273 | 0.007579 | 0.007482 |
| 0.007548 | 0.007530273 | 0.007579 | 0.007482 |
| 0.007515 | 0.007530273 | 0.007579 | 0.007482 |
| 0.007528 | 0.007530273 | 0.007579 | 0.007482 |
| 0.007531 | 0.007530273 | 0.007579 | 0.007482 |
| 0.007552 | 0.007530273 | 0.007579 | 0.007482 |
| 0.007514 | 0.007530273 | 0.007579 | 0.007482 |
| 0.007517 | 0.007530273 | 0.007579 | 0.007482 |
| Sum | 0.082833 | | |
| Mean | 0.007530273 | | |
| SD | 0.00001619 | | |

The process is carried out to obtain the product with standard weight was within the control and the weight was also equivalent to the mean standard weight. Thus optimised results are obtained in the production process by manipulating the variables under consideration so as to enhance the productivity.

## IV. CONCLUSIONS AND LIMITATIONS

The vital reason for the over consumption is the processing settings inputs. It is essentially required to work on the raw material according to its properties to get the best possible outcome. Overheating of material affects the viscosity and hence the flow ratio increases and hence causes loss of material. Under heating causes rejection of finished product due to various defects. Making changes in processing settings not only improves the quality of the product but also brings down the consumption of raw material up to an optimum level.